# A Study of the 2012 January 19 Complex Type II Radio Burst Using Wind, SOHO, and STEREO Observations*

T.B. Teklu, A.V. Gholap

Physics Department, Addis Ababa University
Addis Ababa, Ethiopia
selameykulu@gmail.com

N. Gopalswamy, S. Yashiro, P. Mäkelä, S. Akiyama, N. Thakur and H. Xie

Solar Physics Laboratory, Heliospheric Division
NASA Goddard Space Flight Center
Greenbelt, Maryland, USA
nat.gopalswamy@nasa.gov
Seiji.Yashiro@nasa.gov
pertti.a.makela@nasa.gov
Sachiko.Akiyama@nasa.gov
neeharika.thakur@nasa.gov
hong.xie-1@nasa.gov

*Abstract*—We report on a case study of the complex type II radio burst of 2012 January 19 and its association with a white-light coronal mass ejection (CME). The complexity can be described as the appearance of an additional type II burst component and strong intensity variation. The dynamic spectrum shows a pair of type II bursts with fundamental – harmonic structures, one confined to decameter-hectometric (DH) wavelengths and the other extending to kilometric (km) wavelengths. By comparing the speeds obtained from white-light images with that speed of the shock inferred from the drift rate, we show that the source of the short-lived DH component is near the nose.

*Keywords—coronal mass ejection; type II radio burst; shock.*

## I.  INTRODUCTION

Type II bursts are slowly drifting features in the radio dynamic spectra in contrast to the fast drift of type III bursts. The slow drift is indicative of shock propagation with speeds in the range 500-3000 km s$^{-1}$. The present work utilizes the white-light and radio measurements to analyze the complex type II burst observed by Waves and Plasma Experiment (WAVES, Bourgeret et al. 1995) on board Wind (Acuna et al. 1995) on 2012 January 19 with added input from the Solar and Heliospheric Observatory (SOHO, Domingo et al. 1995) and Solar Terrestrial Relations Observatory (STEREO, Howard et al. 2008) observations. The overall behavior of the type II burst and its association with white-light CME was already reported by Liu et al. (2013). However, they overlooked the presence of a short-duration, DH type II burst at the beginning with a drift rate much higher than that of the long-lasting type II over the same frequency range. In fact, this type II burst is observed by all the three radio receivers in Wind, STEREO Ahead (A), and STEREO Behind (B). We compare the CME nose speed (from height-time measurements) with the speed of the shock estimated from the radio dynamic spectrum. If the two speeds are comparable then the shock-nose should be the source.

## II.  OBSERVATION

The 2012 January 19 type II radio burst is observed by the WAVES instrument on board the STEREO and Wind spacecraft. The STEREO-A and B spacecraft were located at W108° and E113°, respectively on this day. Fig. 1.1 shows the Wind/WAVES dynamic spectrum of a DH type II burst with fundamental (F) – harmonic (H) pair, marked Segment 1 (S1) in addition to the one extending from ~14 MHz (DH) to km wavelengths. The regular type II in the DH-km wavelength is marked Segment 2 (S2), Segment 3 (S3), and Segment 4 (S4). S1 is a short-duration type II burst that lasts for about 10 minutes. This paper is primarily concerned with the presence of S1. S1 is brighter and its slope in the frequency-time plane is steeper. The type II burst is associated with an energetic CME observed by the coronagraphs on board STEREO and Large Angle Spectrometric Coronagraph (LASCO, Brueckner et al. 1995). We refer to this CME as the primary CME (PCME), which first appeared at 14:48 UT on 2012 January 19 in the LASCO/C2 FOV. The CME was reported to appear at 14:36 UT in the Coordinated Data Analysis Workshop (CDAW) CME Catalog (Yashiro et al. 2004; Gopalswamy et al. 2009), but we now know that the time corresponds to the appearance of a preceding CME (CME1), which is quickly overtaken by the PCME. The PCME is associated with an M3.2 flare from National Oceanic and Atmospheric Administration (NOAA) active region (AR) 11402 located at N32E22. A snapshot of the PCME together with preceding CMEs, the associated GOES soft X-ray flare and the STEREO-B/COR1 images are shown in Fig. 1.2. The eruption occurred in the northeast quadrant as can be seen from the difference image at 193 Å from the AIA on aboard the SDO spacecraft overlaid on the LASCO/C2 image.

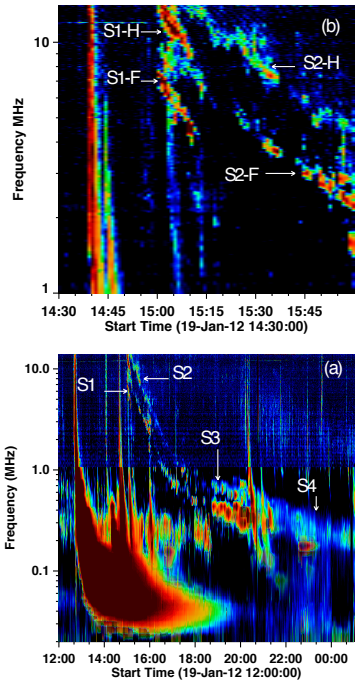

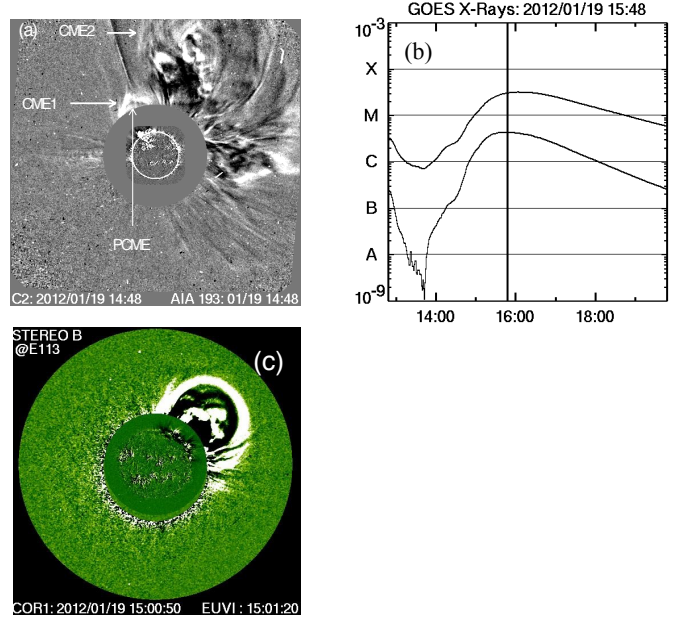

Figure 1.1. (a) The complex dynamic spectrum of the 2012 January 19 type II radio burst observed by WIND/WAVES showing all the segments. The "S1" represents for segment 1. (b) The WAVES dynamic spectrum mainly showing "S1-F" and "S1-H" which represent the fundamental and harmonic components of S1. Others represent for the DH-km range.

Figure 1.2. (a) The white-light running difference image of the CME of interest in the LASCO/C2 FOV at 14:48 UT. The LASCO image is overlaid by a running difference image obtained by the SDO/AIA image at 193 Å. PCME, CME1 and the flare location are indicated in arrows. (b) GOES soft X-ray profiles (upper 1-8 A; lower 0.5-4 A) showing the M3.2 flares associated with PCME. (c) similar to (a) but in STEREO-B/COR1 FOV seen at 15:05 UT.

## III. ANALYSIS AND DISCUSSION

### A. CME Kinematics from White-light Observations

The height-time (H-T) measurements of the PCME from the CDAW CME Catalog shows that the PCME is moving with an average speed of 1120 km s$^{-1}$. It further shows that the PCME has a clear acceleration in the LASCO FOV, somewhat similar to CMEs associated with quiescent filament eruption (Gopalswamy et al. 2015). The acceleration is about 54 m s$^{-2}$, consistent with the large shock formation height indicated by the lack of metric type II radio burst. A close examination of the solar source revealed that a filament immediately to the north of the active region erupted and became part of the CME.

The PCME is a limb event in STEREO-B view, so we are able to track the leading edge using STEREO-B/COR1, COR2 and HI1 images to get a speed that is not subject to projection effects. We are not able to track the PCME in HI2 images because the image quality is too low. The leading edge of the PCME in STEREO-B is at 320° and the corresponding H-T plot is displayed in Fig. 1.3 (a). A second order polynomial is fitted to the COR 1 and COR 2 H-T data. A linear fitted is assumed to the HI1 data. It shows three phases of the PCME: the early acceleration phase followed by the deceleration phase, and finally a constant speed phase (Gopalswamy et al. 2001). The PCME is on average moving with a linear speed of 1255 km s$^{-1}$ at 320°. It has a peak speed of 1584 km s$^{-1}$ at 16:39 UT in the COR2 FOV. Note that this time of peak speed is almost 2 hours after the first appearance of the PCME in the LASCO FOV. When the PCME attained its peak speed, it is at a heliocentric distance of 13.1 Rs. The in-situ shock speed measured by Wind at 1 AU is 408 km s$^{-1}$ (SOHO/MTOF), suggesting a significant deceleration of the PCME beyond the LASCO FOV.

### B. CME Kinematics from Radio and White-light Observations

The speed of the PCME-driven shock can be determined from The speed of the PCME-driven shock can be determined from the drift rate, $V_{sh} = [2L_c/f_c][df/dt]$ where "$f_c$" is the plasma central frequency corresponding to the fundamental component of the type II burst (see e.g., Gopalswamy 2011). "$f_c$" and $df/dt$ can be measured from the dynamic spectrum. The scale height $L_C, \{[1/n][dn/dr]\}^{-1}$, is obtained from the ambient density $n(r)$. If the density decreases with "$r$" as a power law, $n = n_o r^{-\alpha}$, with $n_o$ as the density at the coronal base, one can see that $L_c = r_c/\alpha$. The density profile can also be obtained inverting the fundamental emission frequency ($f = 9.11\sqrt{n}$ kHz) in the dynamic spectrum using heights measured. For each height-time data point of the PCME, the corresponding fundamental emission frequency is noted from the dynamic spectrum. In this section, the speeds of the PCME and shock are compared (see Table 1.1) using "$\alpha$" from power law in Fig. 1.3. The density corresponding to S1 can be fit to LDB distribution with a multiplier of 3.5 (see Fig. 1.3). Here, we have used the basic LDB distribution because the 1-AU density is close to 7.2 $cm^{-3}$. The power law fit yield "α" value of 4.3 for S1 with $n_o = (3-7) \times 10^7$ $cm^{-3}$ (see Fig. 1.3).

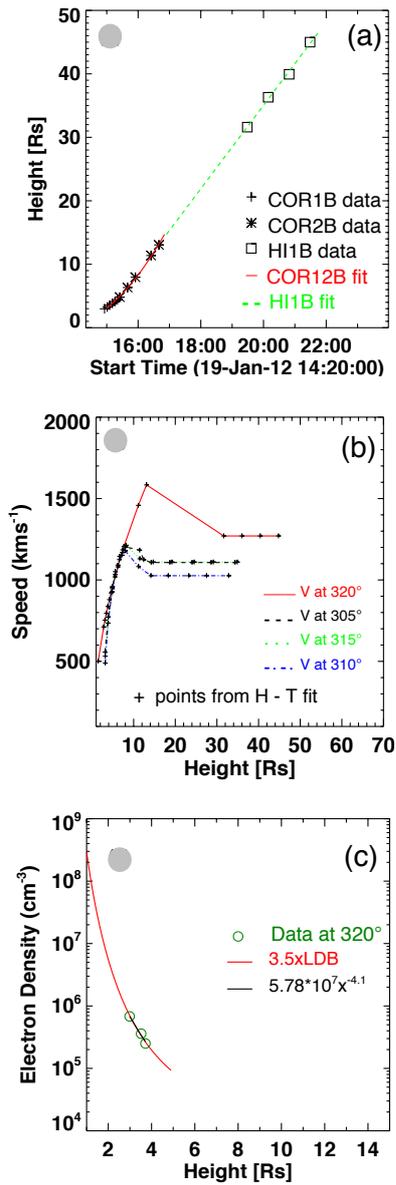

Figure 1.3. (a) height-time measurements of the PCME at the 320°. The "+", "*" and "□"s represent the COR1, COR2 and HI1 measurements, respectively. The solid red curve shows the second order polynomial fit to COR1 and COR2 data and broken green line is a linear fit to the HI1 data. (b) the speed profile of the PCME in (a) at 4 PAs, the red corresponds to the speed profile at 320°. (c) the density profile inferred from type II burst (S1). The green "○" symbol, the red line and the black line indicate the density inverted from the fundamental DH, Leblanc fit and power fit, respectively. "x" is height of PCME from the surface of the Sun.

The derived "α" value is consistent with the typical value expected in the DH wavelength domain (see Gopalswamy 2011). The shock and PCME speeds estimated from radio and white-light observations for S1 are close. The result indicates that the nose region might be the source of S1. Images from STEREO COR1 when compared with the radio dynamic spectrum from S/WAVES shows that the radio burst starts and ends when the PCME is within the COR1 FOV in the height range 3.2–3.7 Rs. Interestingly, this height range corresponds to the peak in the Alfven speed profile of the corona (Mann et al. 1999; Gopalswamy et al. 2001). Therefore, the PCME is barely super-Alfvenic at the time of S1. Note that the PCME is still accelerating around this time and reaches its peak speed an hour later. This means S1 ends mainly because it could not remain superAlfvenic because the speed increase is not fast enough to overcome the increase in Alfven speed with heliocentric distance. The nose region is also consistent with the fact that the burst is seen at all three views (SOHO and STEREO) because the PCME body poses less obstruction to the burst.

Table 1.1 The measured parameters corresponding to S1.

| $f_c$ | $df/dt$ | n | $r_c$ | $L_c$ | $V_{sh}$ | $V_{CME}$ |
|---|---|---|---|---|---|---|
| 5.97 | $4.8 \times 10^{-3}$ | $4.4 \times 10^5$ | 3.45 | 0.8 | 902 | 795 |

$f_c$, $df/dt$, n, α, $r_c$, $L_C$, , $V_s$ and $V_{PCME}$ refer to the average frequency in S1 in MHz, drift rate of S1 in MHz $s^{-1}$, density in $cm^{-3}$, the power law coefficient, CME height (in Rs) at $t_c$, scale height (in Rs) at $r_c$, shock speed in km $s^{-1}$ and primary CME speed in km $s^{-1}$, respectively.

IV. CONCLUSION

The PCME is propagating at a linear speed of 1255 km $s^{-1}$ based on the height-time measurements of the PCME from STEREO-B/COR1, COR2 and HI1. It is driving a strong shock that is responsible for the entire type II emission. However, the shock forms only when the PCME is at a height of ~3 Rs, very similar to filament eruption events. The close speeds of the PCME and shock assuming the density profile follows the power law density distribution, and the simultaneous observation of the segment in the 3 spacecraft, S1 seems to be produced near the nose of the PCME when it is between 3 and 4 Rs. Liu et al. (2013) overlook the details of the radio dynamic spectrum. Basically, they did not consider the two separate F-H structures (S1 and S2).

## Acknowledgments

We thank the Wind/WAVES, SOHO/LASCO and STEREO/SECCHI teams for providing the data. SOHO is an international cooperation project between ESA and NASA.